\documentclass[useAMS,usenatbib]{aa}
\usepackage{graphicx}
\usepackage{txfonts}
\begin{document}
   \title{Hot subdwarfs from the stable Roche lobe overflow channel}


   \author{S. Yu
          \inst{1,2}
          ,
          \and
          L. Li\inst{2}
          }

   \offprints{Shenghua Yu}

   \institute {Armagh Observatory, College Hill, Armagh BT61 9DG,
Northern Ireland, U.K.\\
              \email{syu@arm.ac.uk}
         \and National Astronomical Observatories / Yunnan
Observatory, the Chinese Academy of Sciences, P.O.Box 110, Kunming,
650011, China
             \\
             }



  \abstract
   {Hot subdwarfs are core-helium-burning stars
with extremely thin envelopes. We discuss the formation and
evolution of hot subdwarfs formed through the stable Roche lobe
overflow (RLOF) channel of intermediate-mass binaries, although
their formation channels are various.}
   {In this study, we concentrate on the formation and evolution of hot subdwarfs
binaries through the stable RLOF channel of intermediate-mass
binaries. We aim at setting out the properties of hot subdwarfs and
their progenitors, so that we can understand the formation and
evolution of hot subdwarfs.}
   {Employing Eggleton$'$s stellar
evolution code, we have computed conservative and nonconservative
population I binary evolution sequences. The initial mass of the
primary ranges from 2.2 to 6.3 $M_{\odot}$, spaced by approximately
0.1 in log$M$, the initial mass ratio $q_{\rm i} = M_{1}/M_{2}$ is
between 1.1 and 4.5, and the Roche lobe overflow begins at the main
sequence, the Hertzsprung gap and the first giant branch. In
nonconservative binary evolution, we assume that 50 percent of the
mass lost from the primary leaves the system, carrying away the
specific angular momentum of the primary, and the remaining mass is
accreted on to the secondary during the RLOF. Also, we have studied
the distributions of the mass and orbital periods of hot subdwarfs
using the population synthesis approach.}
   {We have obtained the ranges of the initial parameters of progenitor
binaries and the properties of hot subdwarfs through the stable RLOF
channel of intermediate-mass binaries, e.g. mass, envelope mass and
age of hot subdwarfs. We have found that hot subdwarfs could be
formed through stable Roche lobe overflow at the main sequence and
Hertzsprung gap. We have also found that some subdwarf B or OB stars
have anomalously high mass ($\sim$1 $M_{\odot}$) with a thick
envelope ($\sim$0.07 $M_{\odot}$ $-$ $\sim$0.16) in our models. By
comparing our theoretical results with observations on the hot
subdwarfs in open clusters, we suggest that more hot subdwarfs in
binary systems might be found in open clusters in the future.}
   {}


   \maketitle
%

\section{Introduction}

Hot subdwarfs are generally considered to be core-helium-burning
stars with extremely thin hydrogen-rich or helium-rich envelopes.
Earlier observations (Humason \& Zwicky \cite{Humason47}; Feige
\cite{Feige58}; Greenstein \cite{Greenstein66}; Slettebak et al.
\cite{Slettebak61}; Klemola \cite{Klemola62}; Berger
\cite{Berger63}) showed that a number of faint blue field stars with
early type at high galactic latitudes display peculiar spectra.
Sargent \& Searle (\cite{Sargent68}) defined subdwarf B (sdB) stars
as stars with colors corresponding to those of B stars in which the
Balmer lines are abnormally broad compared to population I main
sequence B stars and they have weaker HeI lines for their color.
Similarly, they defined subdwarf O (sdO) stars as stars that have
strong Balmer lines relative to main sequence stars and in which
HeII $\lambda$4686 is seen. An intermediate class (subdwarf OB, sdOB
stars), which has effective temperature and photospheric helium
abundances between those of sdB stars and sdO stars, has been
reported by Baschek \& Norris (\cite{Baschek75}) and Hunger et al.
(\cite{Hunger81}). As a blueward extension of the horizontal branch
(HB), hot subdwarfs include three subgroups from their
spectroscopic classification (Heber \cite{Heber86}):\\
(i) sdO - display strong HeII or HeI lines. \\
(ii) sdOB - display hydrogen lines and helium lines.\\
(iii) sdB - display strong broadended hydrogen lines and weak

\noindent ~~~~~~~~~~~~~~~~ helium lines.

The origin of hot subdwarfs remains unclear. In the past two
decades, two scenarios for the origin of hot subdwarfs have been
proposed, i.e. a single star model and a binary model. The single
star model was prevalent in the 1990s because of the absence of
confirmed hot subdwarfs binaries. The model assumes that the
envelope of a giant may be striped off by stellar wind near the tip
of the first giant branch (FGB) and leave an almost bare helium
core. If the helium core is ignited, the star may become a single
hot subdwarf (Lee \cite{Lee94}; Yi, Demarque \& Oemler \cite{Yi97};
D$'$Cruz et al. \cite{D$'$Cruz96}). Sweigart (\cite{Sweigart97}) has
studied the evolution of globular-cluster stars and suggested that
helium mixing driven by internal rotation substantially increases
the helium abundance in the envelope. This may result in enhanced
mass loss along the FGB and the formation of a hot subdwarf.
However, the single star model is inconsistent with the latest
observations of hot subdwarfs and the UV upturn of elliptical
galaxies which is a strange phenomenon of the rising flux with
decreasing wavelength between the Lyman limit and 2500 \AA found in
almost all elliptical galaxies. (Yi \& Yoon \cite{Yi04}; Han,
Podsiadlowski \& Lynas-Gray \cite{Han07}). Recent observations
indicate that a quantity of hot subdwarfs are found in binaries.
Maxted et al. (\cite{Maxted01}) argued that more than two-thirds of
their candidates are binaries with short orbital periods from hours
to days, and Napiwotzki et al. (\cite{Napiwotzki04}) have found that
roughly two-fifths of their sample are hot subdwarfs binaries. In
order to explain the formation of hot subdwarfs in binary, Han et
al. (\cite{Han02}) suggested a binary model. In such a model, three
main channels are responsible for the formation of hot subdwarfs
i.e. the common-envelope (CE) ejection channel for hot subdwarf
binaries with short orbital periods (0.05 $\sim$ 40 days), the
stable Roche lobe overflow (RLOF) channel for hot subdwarf binaries
with long orbital periods (0.5 $\sim$ 2000 days), and the double
helium white dwarf (WD) merger channel for single hot subdwarfs. In
the CE ejection channel, the progenitor of a hot subdwarf
experiences dynamical mass transfer on the FGB which would result in
a CE and a spiral-in phase, leaving a very close binary after the
envelope has been ejected. If the helium core of the giant was
ignited, it would become a hot subdwarf in a short-period binary
with a companion of a main sequence or white dwarf. In the stable
RLOF channel, the progenitor of the hot subdwarf would also lose
most of its envelope to produce a hot subdwarf through mass
transfer, but the mass transfer would be stable in contrast to the
CE ejection channel. In the merger channel, two helium white dwarfs
in close binary would be driven together by the orbital angular
momentum loss due to gravitational wave radiation or magnetic
braking effects. When the white dwarfs merge and the merged object
ignites helium, this would produce a single hot subdwarf (Webbink
\cite{Webbink84}; Iben \& Tutukov \cite{Iben86}; Han \cite{Han98}).
The binary model has explained the observations and the UV upturn of
elliptical galaxies (Han, Podsiadlowski \& Lynas-Gray \cite{Han07}).

The majority of field hot subdwarfs found by recent observations
(Maxted et al. \cite{Maxted01}; Morales-Rueda et al.
\cite{Morales-Rueda03}; Napiwotzki et al. \cite{Napiwotzki04};
Edelmann et al. \cite{Edelmann05}; Morales-Rueda et al.
\cite{Morales-Rueda05}) is in short-period binaries, while few hot
subdwarfs in close binaries were found in globular clusters (Moni
Bidin et al. \cite{Monibidin06a,Monibidin06b}). This implies that
there would be different formation channels for field and cluster
hot subdwarfs. Burleigh et al. (\cite{Burleigh99}) suggested a blue
star, which was discovered by Elson et al. (\cite{Elson98}) in the
young cluster NGC 1818 in the Large Magellanic Cloud (LMC), is a
possible hot subdwarf. If hot subdwarfs could exist in very young
clusters, their formation would be a puzzle.

Various channels for the formation of hot subdwarfs have been
studied in the binary model (Han et al. \cite{Han02}), as well as
the stable RLOF channel by which a low-mass giant loses most of its
envelope on the FGB. The giant would leave a degenerate core after
the mass transfer stops. If the mass of the degenerate core is high
enough, it would experience a helium flash and the star would appear
as a core-helium-burning hot subdwarf in a binary. In their model,
binary evolution sequences with the donor mass in the range from 0.8
to 1.9 $M_{\odot}$ with the accreting components of WD stars or
neutron stars (NS) were calculated. In addition, it was assumed that
mass transfer takes place at the FGB, and all the mass lost from the
system carried away its orbital angular momentum of the accreting
component. For the purpose of perfecting the stable RLOF channel and
attaining a comprehensive understanding of the formation of hot
subdwarfs following Han$'$s calculation, we have executed a
computation of evolution sequences of intermediate-mass binaries.

In this study, we have simulated the formation and evolution of hot
subdwarfs through the stable RLOF channel in the conservative and
nonconservative case. We have taken into account the intermediate
mass star filling its Roche lobe at the main sequence (MS), the
Hertzsprung gap (HG) or the first giant branch (FGB). The mass
transfer stops once the radius of the donor is smaller than its
Roche lobe. If the mass of the helium core of the donor (the
primary, $M_{\rm 1}$) is high enough, helium is ignited and the star
becomes a hot subdwarf in a binary in which the other component is
probably a main sequence star or a subgiant star. We have obtained
the properties of the hot subdwarfs and the initial parameters space
of their progenitors, e.g. the effective temperature, the surface
gravity, mass, envelope mass, lifetime of hot subdwarfs phase.
Subsequently, we carried out a Monte Carlo simulation in order to
acquire distributions of properties of hot sudwarfs from the stable
RLOF channel of intermediate-mass binaries and compared these
distributions with observations.

The outline of this paper is as follows. In section 2, we describe
the stellar evolution code adopted in this study. In Section 3, we
present the results. The results are discussed in Section 4 and
summarized in Section 5.


\section{The stellar evolution code and the binary population synthesis code}

\subsection{The stellar evolution code and stellar models}

We use Eggleton$'$s (\cite{Eggleton$'$s71}; \cite{Eggleton$'$s72};
\cite{Eggleton$'$s73}) stellar evolution code, which has been
updated with the latest physics over the last three decades (Han,
Podsiadlowski \& Eggleton \cite{Han94}; Pols et al. \cite{Pols95},
\cite{Pols98}). The code uses a self-adaptive non-Lagrangian mesh
and both convective and semiconvective mixing are treated as a
diffusion process. The stellar structure equations, the mesh
equation and the chemical composition equations are solved
simultaneously.

The current code uses an equation of state that includes pressure
ionization and Coulomb interaction (Pols et al. \cite{Pols95}), the
latest opacity tables derived from Iglesias \& Rogers
(\cite{Iglesias96}) and Alexander \& Ferguson (\cite{Alexander94a},
\cite{Alexander94b}) via quadratic interpolation for $X$ = 0.8, 0.7,
0.5, 0.35, 0.2, 0.1, 0 and for $Y$ $=$ 0.5 $-$ $Z$, 0.2 $-$ $Z$ and
0. Nuclear reaction rates come from Caughlan \& Fowler
(\cite{Caughlan88}) and Caughlan et al. (\cite{Caughlan85}), and
neutrino loss rates are from Itoh et al. (\cite{Itoh89},
\cite{Itoh92}).

We use a typical Population I composition with hydrogen abundance
$X$ = 0.700, helium abundance $Y$ = 0.280 and metallicity $Z$ =
0.020 in our computations. We set $\alpha$ = $\textit{l}$/$H_{\rm
p}$, the ratio of the mixing length to the local pressure
scaleheight is equal to 2. Such a value of $\alpha$ gives a roughly
correct lower main sequence, as determined observationally by
Andersen (\cite{Andersen91}). It also well reproduces the location
of the red giant branch in the Hertzsprung-Russell (HR) diagram for
stars in the Hyades supercluster (Eggen \cite{Eggen85}), as
determined by Bessell et al. (\cite{Bessell89}). A fit to the Sun
also result in $\alpha$ $=$ 2 as the most appropriate choice (Pols
et al. \cite{Pols98}).

Convective overshooting is important for the remnant mass of a
binary after RLOF evolution because the overshooting directly
affects the scale of the nuclear reaction region in a star. In this
paper, we follow the work of Schr$\ddot{\rm o}$der, Pols \& Eggleton
(\cite{Schroder97}) and use an approach based on the stability
criterion itself, the $\delta_{\rm ov}$ prescription, by
incorporating a condition that mixing occurs in a region with
$\nabla_{r} > \nabla_{a} -\delta$, with $\delta$ defined as the
product of a specified constant $\delta_{\rm ov}$, the overshooting
parameter and a conveniently chosen factor that depends only on the
ratio $\zeta$ of the radiation pressure to the gas pressure:

\begin{large}
\begin{equation}
\delta = \delta_{\rm ov}/(2.5+20\zeta+16\zeta^{2}).
\end{equation}
\end{large}
We take $\delta_{\rm ov}$ = 0.12, which leads to overshooting
lengths $\textit{l}_{\textrm{\rm ov}}$ between 0.25 and 0.32$H_{\rm
p}$ for the mass range of 2.5$\sim$7 $M_{\odot}$ with typical
$\delta$ values from 0.05 to 0.04. Such $\delta_{\rm ov}$ best fits
eclipsing binaries (Schr$\ddot{\rm o}$der, Pols \& Eggleton
\cite{Schroder97}, Pols et al. \cite{Pols97}), and here we assume
that the influence of convective overshooting is the same in our
binaries as in the eclipsing binaries.

RLOF and stellar wind are involved in our models. RLOF is treated as
a modification of a surface boundary condition, which is written as:

\begin{large}
\begin{equation}
\frac{\textrm{d}m}{\textrm{d}t}=C\textrm{Max}[0,(\frac{r_{\rm
star}}{r_{\rm lobe}})^{3}] ,
\end{equation}
\end{large}
where d$m$/d$t$ is the mass changing rate of the star, $r_{\rm
star}$ is the radius of the star and $r_{\rm lobe}$ is the radius of
its Roche lobe. In principle, the radius of a primary would equal
the radius of its Roche lobe if the $C$ is large enough. If the $C$
is very large, however, the Roche lobe overflow would not be
continuous. Here we take $C$ = 1000$M_{\odot}$yr$^{-1}$ following
Han et al. (\cite{Han02}), so that RLOF proceeds steadily and the
lobe-filling star overfills its Roche lobe as necessary but never
overfills its lobe by much [typically ($r_{\rm star}/$$r_{\rm lobe}$
$-$ 1) $\lesssim$ 0.001] (Han et al. \cite{Han95}, \cite{Han00},
\cite{Han02}). For stellar wind, we adopt Reimers$'$ wind (Reimers
\cite{Reimers75}) mass-loss law:

\begin{large}
\begin{equation}
\dot{M}_{\rm wind}=4\times10^{-13}\eta RL/M ,
\end{equation}
\end{large}
where we use a coefficient $\eta$ = 1/4 (Carraro et al.
\cite{Carraro96}, Iben \& Renzini \cite{Iben83}, Renzini
\cite{Renzini81}).

Our models include two cases for binary evolution:

\noindent (a) conservative case;

\noindent (b) nonconservative case.

\noindent We assume $-\triangle M_{1}$ is lost from the primary,
$-\beta \triangle M_{1}$ is accreted on to the secondary, then
$-(1-\beta)\triangle M_{1}$ is lost from the system, carrying away
the same specific angular momentum as the center of mass of the
primary. The change of the angular momentum of the system $\triangle
J$ is

\begin{large}
\begin{equation}
\frac{\triangle J}{J}=\frac{(1 - \beta)\triangle
M_{1}M_{2}}{M_{1}(M_{1}+ M_{2})},
\end{equation}
\end{large}
where $J$ is the orbital angular momentum of the system, $M_{1}$ is
the mass of the primary, and $M_{2}$ is the mass of the secondary.
We take the mass transfer efficiency $\beta$ = 1.0 for the
conservative case, and $\beta$ = 0.5 for the nonconservative case,
since the results of Paczy$\acute{\rm n}$ski \& Zi$\acute{\rm
o}$$\l$kowski (\cite{Paczynski67}) and Refsdal, Roth \& Weigert
(\cite{Refsdal74}) indicate that $\beta$ is around 0.5 in the
nonconservative case.

The parameter space for the model grid is three dimensional:

\noindent(i) The range of initial mass of the primaries ($M_{\rm
1i}$):

\noindent ~~~~~log$M_{\rm 1i}$ = 0.35 (2.23$M_{\rm \odot}$) $\sim$
0.80 (6.31$M_{\rm \odot}$), and $\bigtriangleup$log$M_{\rm 1i}$ =
0.05;

\noindent(ii) The initial mass ratio $q_{\rm i}=M_{1}/M_{2}$:

\noindent     ~~~~~~$q_{\rm i}$ = 1.1, 1.5, 2.0, 3.0, 4.0, 4.2, 4.5;

\noindent(iii) The initial orbital periods of the binaries from the
minimum

\noindent ~~~~~~ period, at which a zero-age main sequence (ZAMS)
star

\noindent ~~~~~~ would fill its Roche lobe, to the maximum period,
at which

\noindent ~~~~~~ a star would fill its Roche lobe at the FGB.

In this paper, we focus just on the stable RLOF channel and do not
discuss the case of dynamically unstable mass transfer. In each
calculation, we checked whether mass transfer was dynamically
stable. If the mass transfer rate keeps increasing rapidly,
dynamical instability would occur as discussed by Podsiadlowski et
al. (\cite{Podsiadlowski02}). This leads to the stellar evolution
code breaking down. If mass transfer is stable, we continue mass
transfer until the donor shrinks below its Roche lobe, terminating
the mass transfer phase. In the case of stable mass transfer, if the
mass of the helium core exceeds the appropriate minimum core mass
for helium ignition, it will appear as a core-helium-burning hot
subdwarf.

\subsection{Monte Carlo simulation parameters}

In order to obtain distributions of properties of hot subdwarfs from
the stable RLOF channel of intermediate-mass binaries, we have
performed a simple Monte Carlo simulation where we follow the
evolution of a sample of a million binaries. The physical inputs of
the simulation are depicted as follows:

(i) We assume that the star formation rate (SFR) is constant over
the last 13.5 Gyr.

(ii) For the initial mass function (IMF) of the primary, we adopted
the IMF of Miller \& Scalo (\cite{Miller79}); the primary mass is
generated using the formula of Eggleton, Fitchett \& Tout
(\cite{Eggleton$'$s89}),

\begin{large}
\begin{equation}
M_{1}=\frac{0.19X}{(1-X)^{0.75}+0.032(1-X)^{1/4}},
\end{equation}
\end{large}
where $X$ is a random number uniformly distributed between 0 and 1.
The studies by Kroupa, Tout \& Gilmore (\cite{Kroupa93}) and Zoccali
et al. (\cite{Zoccali00}) support this IMF.

(iii) For the initial mass ratio distribution, we suppose a constant
one (Mazeh et al. \cite{Mazeh92}; Goldberg \& Mazeh
\cite{Goldberg94}),

\begin{large}
\begin{equation}
n(1/q) = 1,  0 \leqslant 1/q \leqslant 1,
\end{equation}
\end{large}
where $q_{\rm i} = M_{1}/M_{2}$.

(iv) For the distribution of initial orbital separations, we
employed that used by Han et al. (\cite{Han03}), where they assume
that all stars are members of binary systems and that the
distribution of separations is constant in log$a$ ($a$ is the
separation) for wide binaries and falls off smoothly at close
separations:

\begin{large}
\begin{equation}
an(a)=\left\{
\begin{array}{c}
\alpha_{\rm sep}(\frac{a}{a_{0}})^{m},a\leqslant
a_{0},\\
\alpha_{\rm sep},a_{0}<a<a_{1}.
\end{array}
\right.
\end{equation}
\end{large}

where $\alpha_{\rm sep}$$\approx$0.070, $a_{0}=10R_{\odot}$,
$a_{1}=5.75\times 10^{6}R_{\odot}=0.13$ pc, $m\approx1.2$. This
distribution implies that there is an equal number of wide binary
systems per logarithmic interval, and that approximately 50 per cent
of stellar systems are binary systems with orbital periods less than
100 yr.

The total numbers of observed hot subdwarfs are associated with
their evolution lifetime. Each binary system in our models is
multiplied by a factor of $\gamma_{ \rm t}$, where $\gamma_{\rm
t}=t/2\times10^{8}$ yrs, $t$ is the lifetime of hot subdwarf phase.

%
\section{Results}

\begin{figure*}
\centering
\includegraphics[width=12cm,clip]{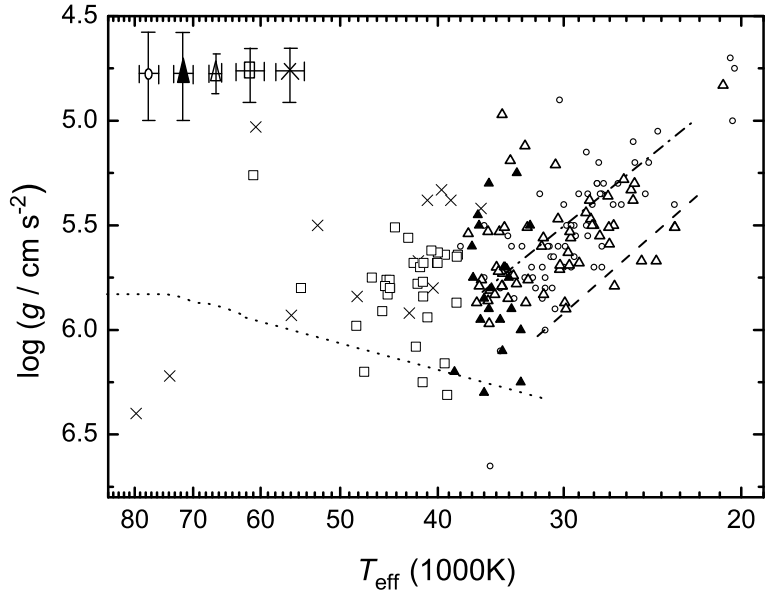}
\caption{Logarithmic surface gravities versus effective temperature
of observed hot subdwarfs (see typical error-bars in the upper left
side). Open circles and solid triangles are for sdB and sdOB stars
observed by Edelmann et al. (\cite{Edelmann03}). Open triangles are
for sdB stars observed by Lisker et al. (\cite{Lisker05}). Crosses
are for sdO stars observed by Stroeer et al. (\cite{Stroeer07}).
Dashed line, dot-dashed line and dotted line indicate ZAEHB (the
Zero-Age Extreme Horizontal Branch, assuming the mass of hot
subdwarf is 0.5 $M_{\odot}$.), TAEHB (the Terminal-Age Extreme
Horizontal Branch, assuming the mass of hot subdwarf is 0.5
$M_{\odot}$.) (Dorman et al. \cite{Dorman93}) and He-ZAMS (the
Helium Zero-Age Main Sequence) (Paczy$\acute{\rm n}$ski
\cite{Paczynski71}).}
\end{figure*}

In general, three subgroups are defined from photometric and
spectroscopic classifications: sdB, sdOB and sdO. Fig.1 displays the
position of observed hot subdwarfs. It is seen in Fig.1 that there
is a boundary between sdB, sdOB and sdO stars at effective
temperature $T_{\rm eff} \approx $35,000 K. The effective
temperature of sdB and sdOB stars is between 20,000 K and 35,000 K
while that of sdO is between 35,000 K and 80,000 K. Due to the tight
correlation of spectral type (effective temperature, $T_{\rm eff}$)
and spectroscopy, we define three subgroups from spectral type
($T_{\rm eff}$) classification corresponding to the spectroscopic
classification:

(1) sdB - $T_{\rm eff}$ is between 20,000 K $\sim$ 35,000 K.

(2) sdOB - $T_{\rm eff}$ is between 35,000 K $\sim$ 40,000 K.

(3) sdO - $T_{\rm eff}$ is between 40,000 K $\sim$ 80,000 K.
\\All of them have helium-burning-cores, and their logarithmic surface
gravities (log($g$)) are between 4.5 and 6.5 (cgs). We distinguish
sdB and sdOB stars by effective temperature for discussion, although
their differences are also associated with surface helium abundance.
In fact, sdOB and sdB stars are mixed in the region 34,000 - 38,000
K in the $T_{\rm eff}$-log($g$) diagram.

\subsection{The initial parameter space}

We are interested in the ranges of the initial parameters of the
progenitors of hot subdwarfs through the stable RLOF channel of
intermediate-mass binaries, i.e. the primary mass ($M_{\rm i}$), the
mass ratio ($q_{\rm i}$), the orbital periods (log$P$$_{\rm i}$).
Figs. 2 and 3 show the initial parameter space of the progenitors of
hot subdwarfs in conservative case and nonconservative case.

From Figs. 2 and 3, we see that the number of binary systems
experiencing stable mass transfer decreases with increasing mass
ratio $q_{\rm i}$. When $q_{\rm i}>$3.0, a large number of binary
systems in the conservative case have mass transfer rates higher
than 10$^{-5}$ $M_{\odot}$yr$^{-1}$ (the light grey region). The
number of binaries with stable mass transfer in the nonconservative
case is more than that in the conservative case due to mass loss
from binaries. The primary lying in the light grey region in Figs. 2
and 3 would become a hot subdwarf in $\sim$10$^{6}$yrs after it
experiences a stable mass transfer with a rate higher than 10$^{-5}$
$M_{\odot}$yr$^{-1}$ which last $\sim$10$^{5}$ to $\sim$10$^{6}$
yrs. Given $q_{\rm i}>$4.2 in the conservative case ($q_{\rm i}>$4.5
in the nonconservative case), the mass transfer of all systems in
our models is dynamically unstable, and the stellar evolution code
breaks down.

The formation of different types of hot subdwarfs depends greatly on
the initial parameters (i.e. initial mass, initial orbital periods)
of binaries, although this relationship becomes weaker if $q_{\rm
i}\geq$4.0. We depict these relationships as follows:

(i) We can understand the relationship between the formation of
different types of hot subdwarfs and the initial mass of their
progenitors. A star with a larger initial mass could leave a larger
helium core. If the helium core is ignited, the star becomes a hot
subdwarf with an earlier spectral type.

(ii) The relationship between the formation of different types of
hot subdwarfs and the initial orbital periods is more complicated
than the relationship mentioned above.

From Figs. 2 and 3, for a given initial mass, such as $M_{\rm
i}=$5.01$M_{\odot}$, we can see that, if $q_{\rm i}\leq$3.0, it may
produce sdB stars (sdOB or sdO stars) when the initial orbital
periods are shorter than 2.1 days (longer than 2.1 days); if $q_{\rm
i}>$3.0, only sdO stars will be produced. This is because an MS star
fills its Roche lobe and this results in the onset of the mass
transfer at the MS stage. In this case, a star loses more mass
because of a longer mass transfer time, so that a smaller helium
core and a thinner envelope would be left, when the star enters the
extreme horizontal branch (EHB) stage.

In addition, if $q_{\rm i}\leq$1.5, for a given initial mass, such
as $M_{\rm i}=$5.01$M_{\odot}$, we can see that sdB or sdOB stars
(sdO stars) may also be produced when the initial orbital period is
longer than 31.6 days (shorter than 31.6 days but longer than 2.1
days), where the mass transfer begins at the FGB stage. In this
case, the timescale of the mass transfer is very short, as the
primary contracts soon. This leads to less mass lost from the
primary, leaving an sdB star with a larger helium core and thicker
envelope.

\begin{figure*}
\centering
\includegraphics[width=8cm,clip,angle=270]{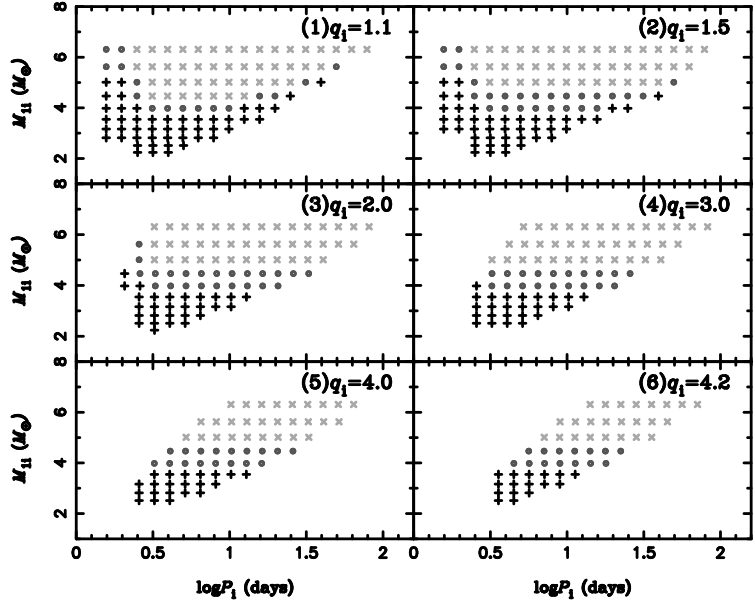}
\caption{Initial parameters of progenitors of hot subdwarfs in the
conservative case. Pluses (black) are for progenitors of sdB stars,
circles (dark grey) for progenitors of sdOB stars, and crosses
(light grey) for progenitors of sdO stars. Panels (1) $\sim$ (6)
indicate the mass ratio $q_{\rm i}$ of 1.1, 1.5, 2.0, 3.0, 4.0, 4.2,
respectively.}
\end{figure*}

\begin{figure*}
\centering
\includegraphics[width=8cm,clip,angle=270]{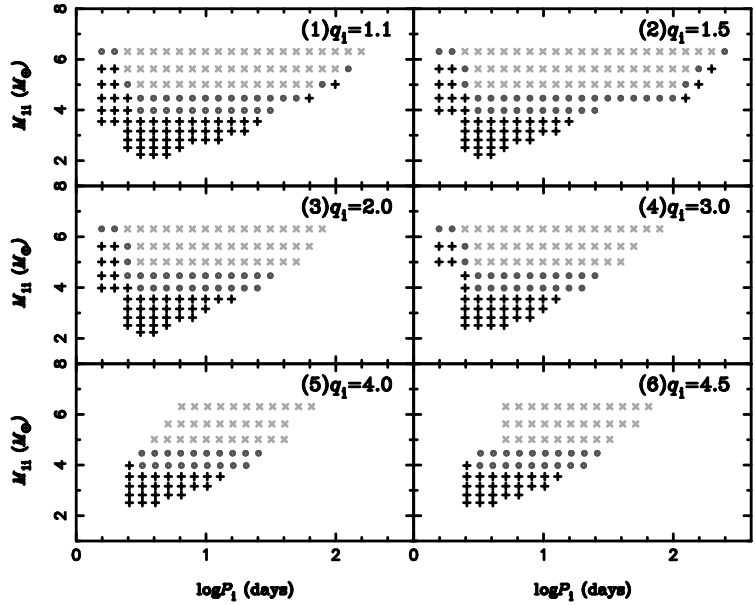}
\caption{Initial parameters of progenitors of hot subdwarfs in the
nonconservative case. Pluses (black) are for progenitors of sdB
stars, circles (dark grey) for progenitors of sdOB stars, and
crosses (light grey) for progenitors of sdO stars. Panels (1) $\sim$
(6) indicate the mass ratio $q_{\rm i}$ of 1.1, 1.5, 2.0, 3.0, 4.0,
4.5, respectively.}
\end{figure*}

\subsection{The $T_{\rm eff}$-log($g$) diagram}

The evolution tracks of some hot subdwarfs with different initial
parameters are shown in Figure 4. Panels (a) and (b) are for the
conservative case while Panels (c) and (d) for the nonconservative
case.

As seen from Fig. 4, the effective temperatures $T_{\rm eff}$ of hot
subdwarfs increase with increasing initial mass while their
logarithmic surface gravities log($g$) decrease. This is because a
star with higher initial mass could more easily leave a larger
helium core. For a given initial mass, log($g$) of hot subdwarfs
decreases with increasing initial orbital periods.

We find that the relationship between $T_{\rm eff}$ and $P$$_{\rm
i}$ is somewhat strange both in the conservative and nonconservative
case, especially for given $q_{\rm i}\leq$1.5 and $M_{\rm
i}\geq$4.0$M_{\odot}$. As mentioned in Section 3.1, this phenomenon
is associated with the mass of the envelope of hot subdwarfs. In
order to display the relationship between $T_{\rm eff}$ and
$P$$_{\rm i}$ clearly, we plot in Fig. 5 the evolution tracks of hot
subdwarfs with different initial mass and orbital periods in the
$T_{\rm eff}$-log($g$) diagram when $q_{\rm i}=$1.5. For a given
initial mass, such as 5.01 $M_{\odot}$, $T_{\rm eff}$ would increase
and then it would decline with increasing $P$$_{\rm i}$.

From Fig. 4, we can see that there is a lack of observed hot
subdwarfs in the upper left region of the $T_{\rm eff}$-log($g$)
diagram. This is consistent with our models (Note that the evolution
curves in Fig. 4 are at the boundary for producing hot subdwarfs.).
In addition, if the mass of the helium core of a star is too small
to be ignited, it will not experience the core-helium-burning stage.
So, it is difficult to observe hot subdwarfs in the lower right
region of the $T_{\rm eff}$-log($g$) diagram.

The evolutionary tracks of hot subdwarfs in our models can cover
positions of observed hot subdwarfs except for a few observed hot
subdwarfs with effective temperatures between 39,000K and 50,000K
and logarithmic surface gravities between 6.1 and 6.4 (cgs). The
origin of these hot subdwarfs is not clear.

\begin{figure*}
\centering
\includegraphics[width=10cm,clip,angle=270]{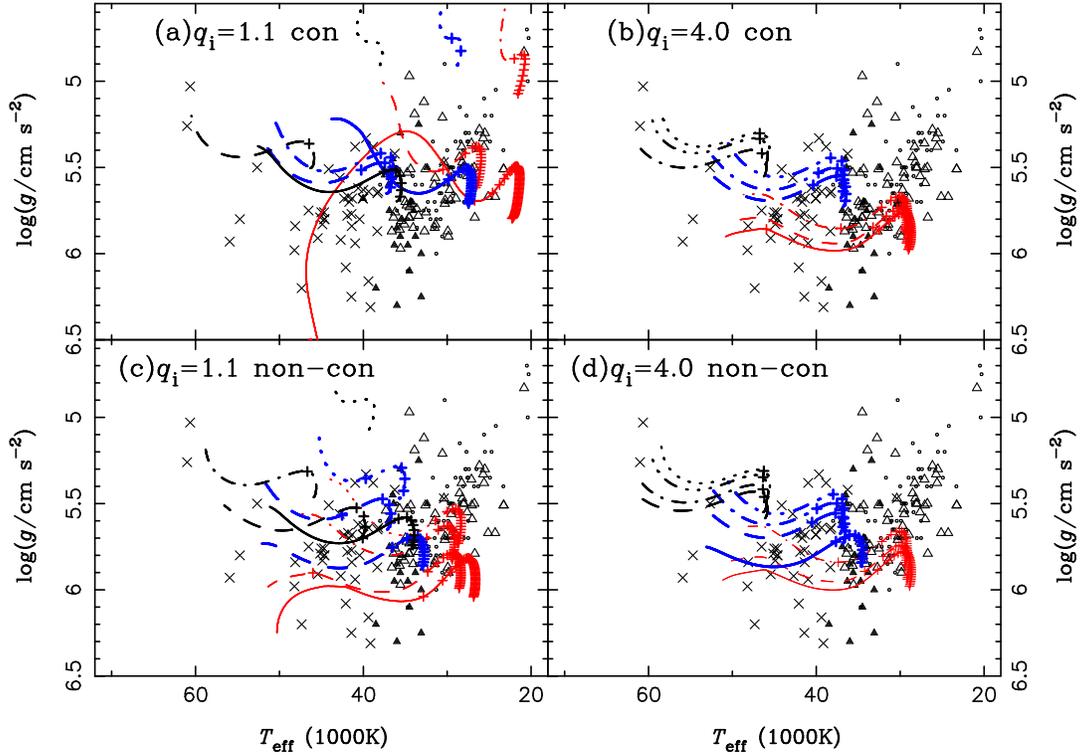}
\caption{Evolutionary tracks of hot subdwarfs with different initial
mass ratios in the $T_{\rm eff}$-log$ g$ diagram. Panels (a)
($q=$1.1) and (b) ($q=$4.0) are for the conservative case. Panels
(c) ($q=$1.1) and (d) ($q=$4.0) are for the nonconservative case.
Circles and triangles are for sdB stars from the Hamburg quasar
survey (Edelmann et al. \cite{Edelmann03}) and the ESO supernova Ia
progenitor survey (Lisker et al. \cite{Lisker05}), respectively.
Squares are for sdO stars from the ESO supernova Ia progenitor
survey (Stroeer et al. \cite{Stroeer07}). Red curves, blue curves
and black curves are for the initial primary mass of
2.82$M_{\odot}$, 3.98$M_{\odot}$ and 5.62$M_{\odot}$. In panel (a),
solid curves are for initial orbital periods of 1.6 days, dashed for
5.0 days, dot-dashed for 7.9 days, blue dotted curves for 20.0 days,
and black dotted curves for 31.6 days. In panel (b), solid curves
are for initial orbital periods of 2.6 days, dashed for 3.2 days,
dot-dashed for 6.5 days, dash-dot-dot-dot for 20.0 days, dotted for
50.1 days. In panel (c), solid curves are for initial orbital
periods of 1.6 days, dashed for 2.5 days, dot-dashed for 6.5 days,
red dotted for 12.6 days, blue dotted for 31.5 days, and black
dotted for 48.5 days. In panel (d), solid curves are for initial
orbital periods of 2.5 days, red and blue dashed for 3.2 days, black
dashed for 5.1 days, dot-dashed for 8.1 days, dash-dot-dot-dot for
20.4 days, dotted for 32.4 days. The age differences between
adjacent crosses are 10$^{7}$ yrs.}
\end{figure*}

\begin{figure*}
\centering
\includegraphics[width=8cm,clip,angle=270]{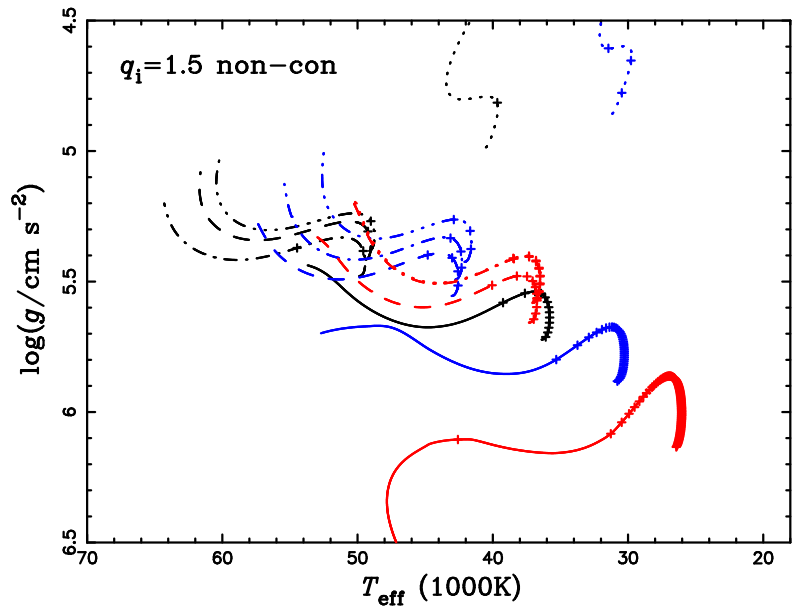}
\caption{Evolutionary tracks of hot subdwarfs with mass ratio
$q=$1.5 in the $T_{\rm eff}$-log$ g$ diagram for the nonconservative
case. Red curves, blue curves and black curves are for the initial
primary mass of 3.98$M_{\odot}$, 5.01$M_{\odot}$ and
6.31$M_{\odot}$. Solid curves are for initial orbital periods of 1.6
days, dashed for 6.3 days, dot-dashed for 24.9 days,
dash-dot-dot-dot for 99.1 days, blue dotted for 157.0 days and black
dotted for 248.9 days. The age differences between adjacent crosses
are 5.0$\times$10$^{6}$ yrs.}
\end{figure*}

\subsection{The mass, the envelope mass and the orbital periods of hot subdwarfs}

We have obtained the mass and the envelope mass of hot subdwarfs
which depend strongly on the initial parameters of their
progenitors, i.e. the primary mass ($M_{\rm 1i}$), the orbital
periods ($P_{\rm i}$) and the mass ratio ($q_{\rm i}$). The mass and
the envelope mass of hot subdwarfs for the conservative case are
shown in Figs. 6 and 7, and they are shown in Figs. 8 and 9 for the
nonconservative case.

The relationship between the envelope mass of hot subdwarfs and the
initial parameters of their progenitors can help us to understand
the post-evolution of hot subdwarfs. As shown by Figs. 7 and 9, the
envelope mass has a weak dependence on the initial mass ratio if
$q_{\rm i}\geq$2.0 while the envelope mass has a clearly wider range
if $q_{\rm i}\leq$1.5. For a given initial mass and initial mass
ratio, the longer the initial orbital periods, the higher the
envelope mass. Hot subdwarfs would enter asymptotic giant branch
manqu$\acute{\rm e}$ (AGB-manqu$\acute{\rm e}$) stage after their
core helium exhaustion if their envelope masses are between
$\sim$0.01 and 0.04 $M_{\odot}$; hot subdwarfs with an envelope mass
of $\sim$ 0.04 $\sim$ 0.08 $M_{\odot}$ would undergo the early
asymptotic giant branch (early-AGB) stage; hot subdwarfs with the
envelope mass larger than $\sim$0.08 $M_{\odot}$ would become normal
asymptotic giant branch (AGB) stars (see Fig. 10 for an example).

As seen from Figs. 6-9, a few sdB stars have progenitors with a mass
of 3.98 - 5.01 $M_{\odot}$ if $q_{\rm i}\leq$3.0, while the majority
of sdB stars come from stars with mass of 2.23 - 3.55 $M_{\odot}$.
This is because a primary with a high initial mass would lose more
mass after a long mass transfer phase due to the primary filling its
Roche lobe at the MS stage. The progenitors of sdO (sdOB) stars have
high mass in general, e.g. $\geq$5.01 $M_{\odot}$ (3.98 $M_{\odot}$
$\sim$ 4.47 $M_{\odot}$).

In particular, we find that if the mass ratio $q_{\rm i}\leq$1.5,
there are some sdB or sdOB stars with anomalously high mass and
envelope mass ($'$anomalous$'$ sdB or sdOB stars, see Table 2.)
formed though the stable RLOF channel, whose progenitors have a high
initial mass and a long initial orbital period, such as $M_{\rm
1i}=$ 5.01 $M_{\odot}$, $P_{\rm i}=$ 40.6 days (see Table 3.). These
results are consistent with the initial-final mass relation
discussed by Han et al. (\cite{Han00}), Chen et al. (\cite{Chen02,
Chen03}).

In order to obtain the mass distribution of hot subdwarfs and the
final orbital period distribution of the binaries in this work, we
have plotted Fig. 11 and 12 by executing a simple Monte Carlo
simulation as mentioned in section 2.2. It is seen that the mass of
hot subdwarfs are in a wide range from 0.33 to 1.22 $M_{\odot}$ with
a peak near 0.45 $M_{\odot}$. The orbital periods of the binaries
are also in a wide range from 5.0 to 900.0 days, and peak at around
120.0 days in the conservative case while the peak at around 30.0
days in the nonconservative case.

In Fig. 12, we have also plotted the orbital periods of observed
binary systems containing hot subdwarfs (Morales-Rueda et al.
\cite{Morales-Rueda03}, Edelmann et al. \cite{Edelmann05},
Napiwotzki et al. \cite{Napiwotzki04}) as short bars with different
colors. A few observed binaries with long orbital periods could be
formed through the stable RLOF channel while the majority of
observed binaries with short orbital periods probably would come
from the CE ejection channel (Han et al. \cite{Han02}).

\begin{table}
\begin{minipage}[t]{\columnwidth}
\caption{The mass and the envelope mass of normal hot subdwarf
stars.}\footnotetext{$M_{\rm f} =$ the mass of hot subdwarfs; $M_{\rm
env} =$ the envelope mass of hot subdwarfs.}
\centering
\renewcommand{\footnoterule}{}  
\begin{tabular}{lccccc}
\hline \hline
 type  &  $M_{\rm f}$  &  $M_{\rm env}$ \\
 & ($M_{\odot}$) & ($M_{\odot}$) \\
\hline
   conservative case:  \\
sdB   &  0.33$\sim$0.67  &  0.01$\sim$0.05 \\  
sdOB  &  0.61$\sim$0.86  &  0.05$\sim$0.10 \\  
sdO   &  0.75$\sim$1.44  &  0.07$\sim$0.22 \\  
\hline
   nonconservative case:  \\
sdB   &  0.32$\sim$0.61  &  0.01$\sim$0.05 \\  
sdOB  &  0.60$\sim$0.85  &  0.05$\sim$0.09 \\ 
sdO   &  0.81$\sim$1.36  &  0.08$\sim$0.19  \\
\hline
\end{tabular}
\end{minipage}
\end{table}

\begin{table}
\begin{minipage}[t]{\columnwidth}
\caption{The mass and the envelope mass of anomalous hot subdwarfs
for mass ratios $q_{\rm i}=$1.1, 1.5.} \footnotetext{$M_{\rm f} =$ the
mass of hot subdwarfs; $M_{\rm env} =$ the envelope mass of hot
subdwarfs.} \label{catalog} \centering
\renewcommand{\footnoterule}{}  
\begin{tabular}{lccccc}
\hline \hline
 type  &  $M_{\rm f}$  &  $M_{\rm env}$ \\
 & ($M_{\odot}$) & ($M_{\odot}$) \\
\hline
   conservative case:  \\
sdB   &  0.72$\sim$1.06  &  0.07$\sim$0.16 \\  
sdOB  &  0.98$\sim$1.23  &  0.11$\sim$0.19 \\  
\hline
   nonconservative case:  \\
sdB   &  0.83$\sim$1.13  &  0.10$\sim$0.16 \\  
sdOB  &  0.97$\sim$1.17  &  0.11$\sim$0.16 \\ 
\hline
\end{tabular}
\end{minipage}
\end{table}

\begin{figure*}
\centering
\includegraphics[width=8cm,clip,angle=270]{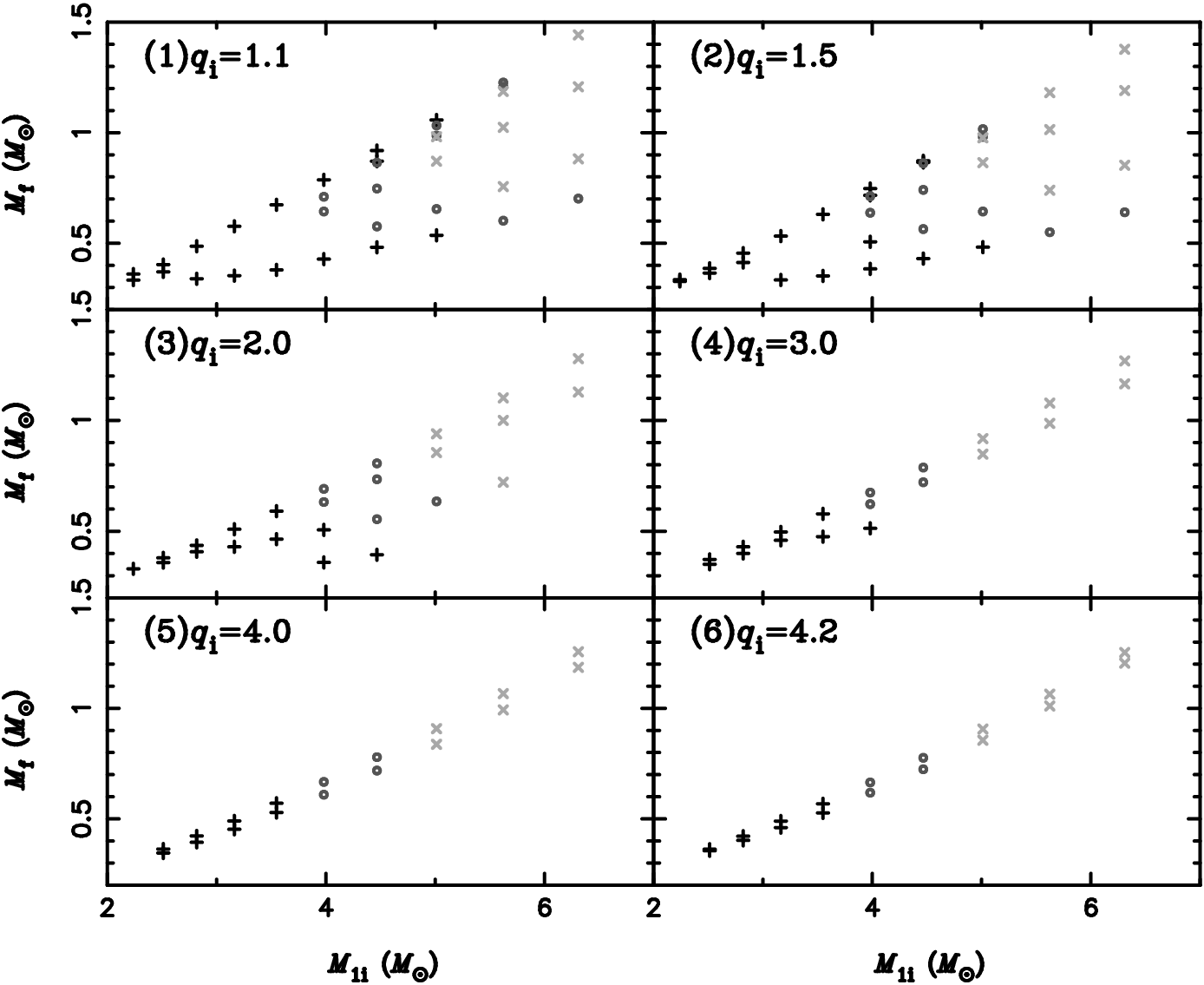}
\caption{The mass of hot subdwarfs ($M_{\rm f}$) versus the mass of
their pogenitors ($M_{\rm 1i}$) in the conservative case. Pluses
(black) are for sdB stars, circles (dark grey) for sdOB stars, and
crosses (light grey) for sdO stars. Panels (1) $\sim$ (6) indicate
the mass ratio $q_{\rm i}$ of 1.1, 1.5, 2.0, 3.0, 4.0, 4.2,
respectively.}
\end{figure*}

\begin{figure*}
\centering
\includegraphics[width=8cm,clip,angle=270]{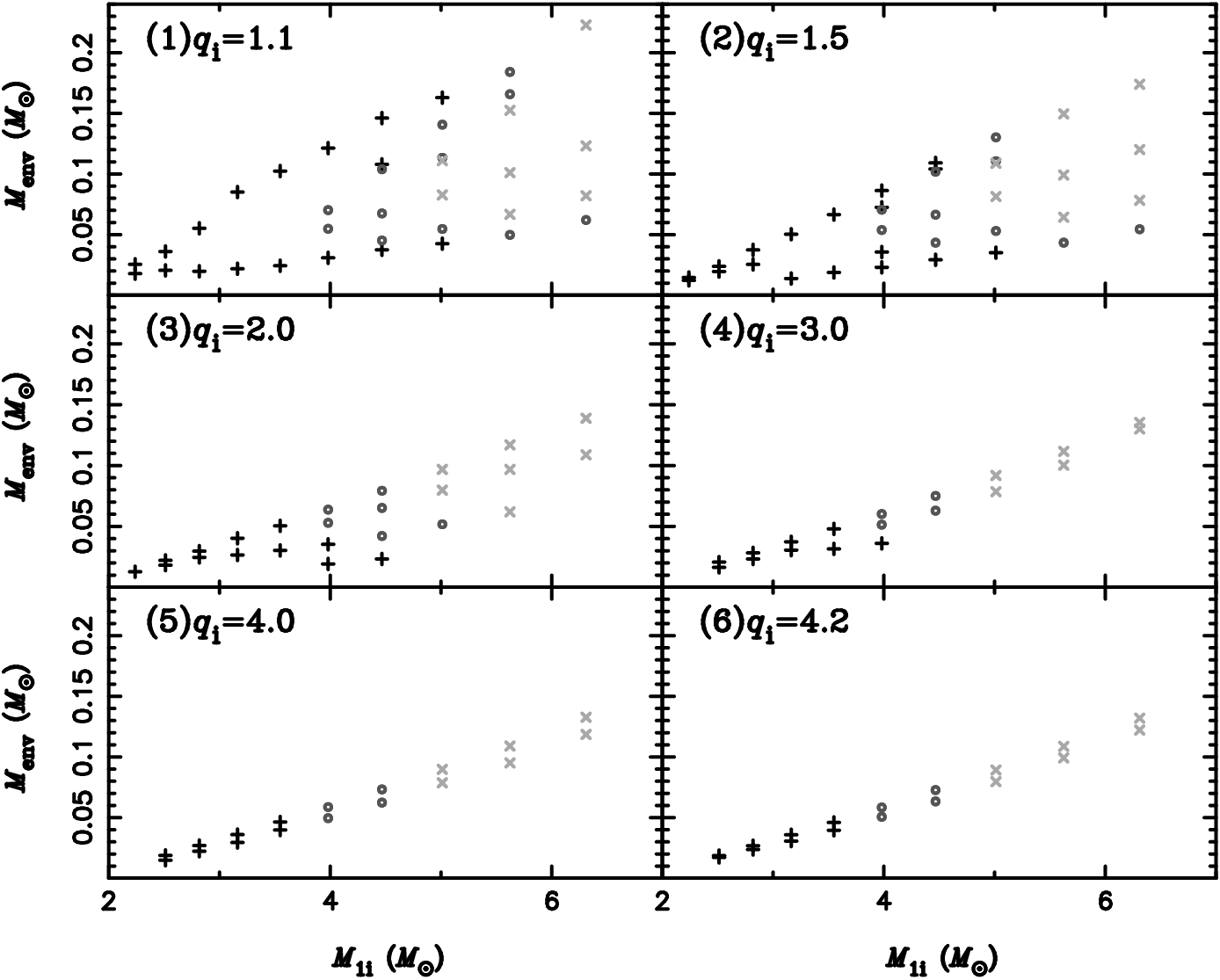}
\caption{The envelope mass of hot subdwarfs ($M_{\rm env}$) versus
the mass of their pogenitors ($M_{\rm 1i}$) in the conservative
case. Pluses (black) are for sdB stars, circles (dark grey) for sdOB
stars, and crosses (light grey) for sdO stars. Panels (1) $\sim$ (6)
indicate the mass ratio $q_{\rm i}$ of 1.1, 1.5, 2.0, 3.0, 4.0, 4.2,
respectively.}
\end{figure*}

\begin{figure*}
\centering
\includegraphics[width=8cm,clip,angle=270]{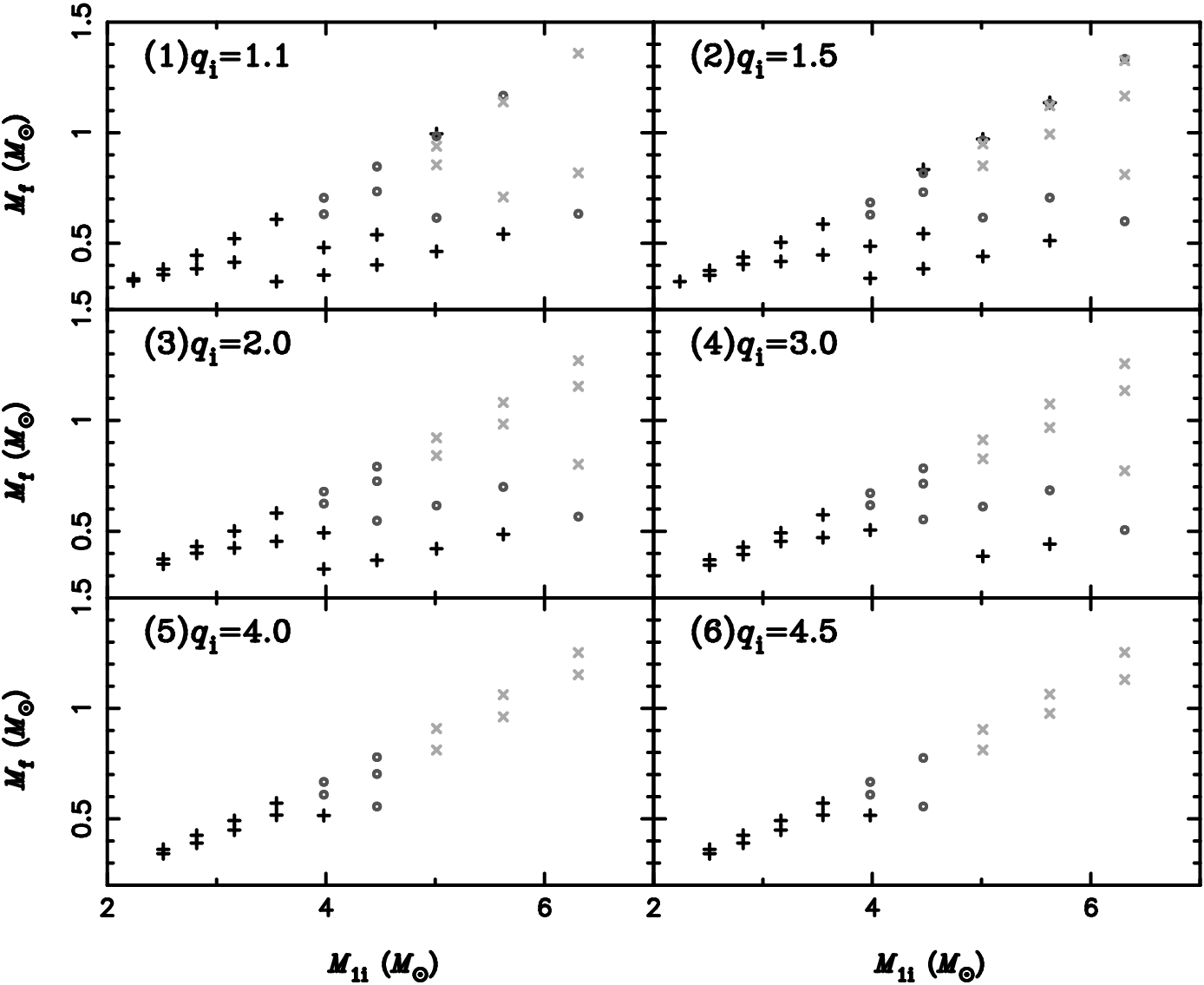}
\caption{The mass of hot subdwarfs ($M_{\rm f}$) versus the mass of
their pogenitors ($M_{\rm 1i}$) in the nonconservative case. Pluses
(black) are for sdB stars, circles (dark grey) for sdOB stars, and
crosses (light grey) for sdO stars. Panels (1) $\sim$ (6) indicate
the mass ratio $q_{\rm i}$ of 1.1, 1.5, 2.0, 3.0, 4.0, 4.5,
respectively.}
\end{figure*}

\begin{figure*}
\centering
\includegraphics[width=8cm,clip,angle=270]{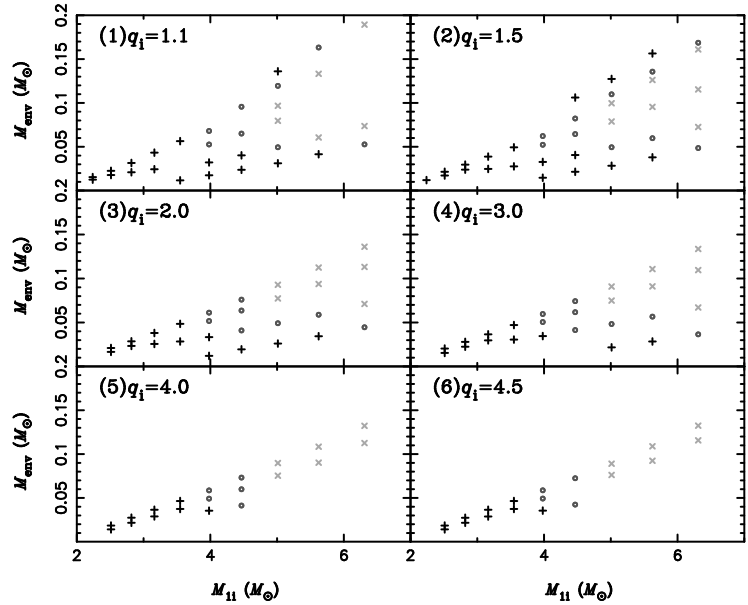}
\caption{The envelope mass of hot subdwarfs ($M_{\rm env}$) versus
the mass of their pogenitors ($M_{\rm 1i}$) in the nonconservative
case. Pluses (black) are for sdB stars, circles (dark grey) for sdOB
stars, and crosses (light grey) for sdO stars. Panels (1) $\sim$ (6)
indicate the mass ratio $q_{\rm i}$ of 1.1, 1.5, 2.0, 3.0, 4.0, 4.5,
respectively.}
\end{figure*}

\begin{figure*}
\centering
\includegraphics[width=8cm,clip,angle=270]{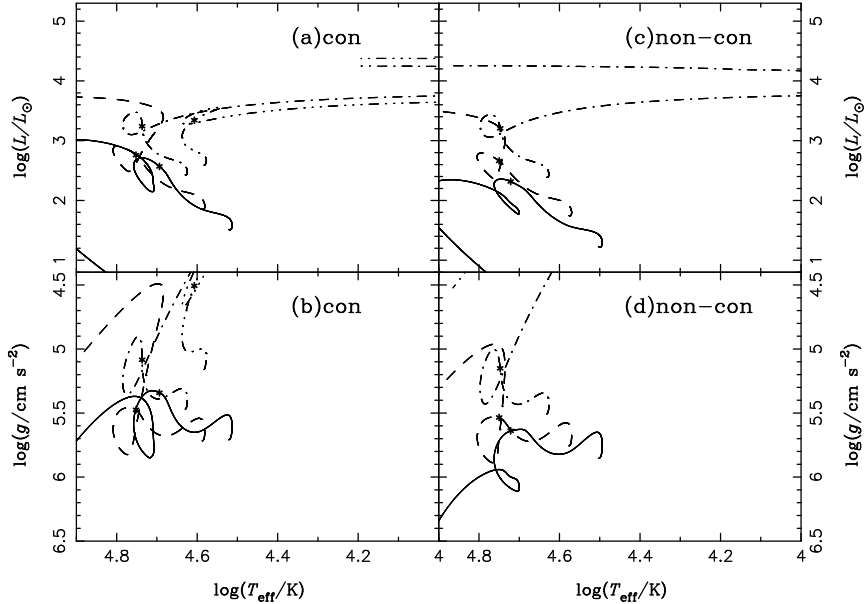}
\caption{The HR diagrams and the $T_{\rm eff}$-log($g$) diagrams of
hot subdwarfs where the initial mass ($M_{\rm 1i}$) is 5.01
$M_{\odot}$ and the mass ratio $q_{\rm i}$ is 1.1. Solid curves are
for initial orbital periods of 1.6 days (sdB stars with envelope
mass of 0.04 $M_{\odot}$ in left panel, envelope mass of 0.03
$M_{\odot}$ in right panel), dashed for 2.5 days (sdOB stars with
envelope mass of 0.05 $M_{\odot}$), dot-dashed for 15.7 days (sdO
stars with envelope mass of 0.09 $M_{\odot}$), dash-dot-dot-dot for
31.3 days (left panel, sdOB star with envelope mass of 0.12
$M_{\odot}$). Panels (a) and (b) are for the conservative case.
Panels (c) and (d) are for the nonconservative case. Asterisks
indicate that
helium is exhausted in its center.}
\end{figure*}

\begin{figure*}
\centering
\includegraphics[width=8cm,clip,angle=0]{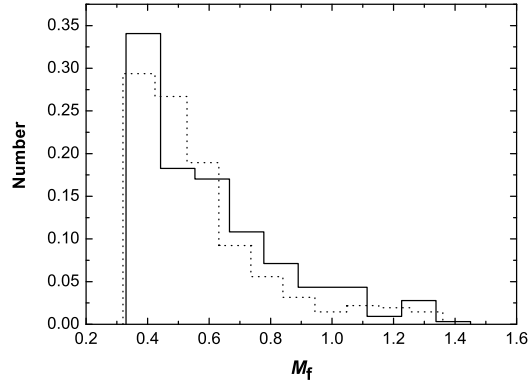}
\caption{The distributions of the masses of hot subdwarfs from the
stable RLOF channel in this paper. $M_{\rm f}$ is the mass of hot
subdwarfs. The solid and dotted line indicate the conservative and
nonconservative cases, respectively.}
\end{figure*}

\begin{figure*}
\centering
\includegraphics[width=8cm,clip,angle=0]{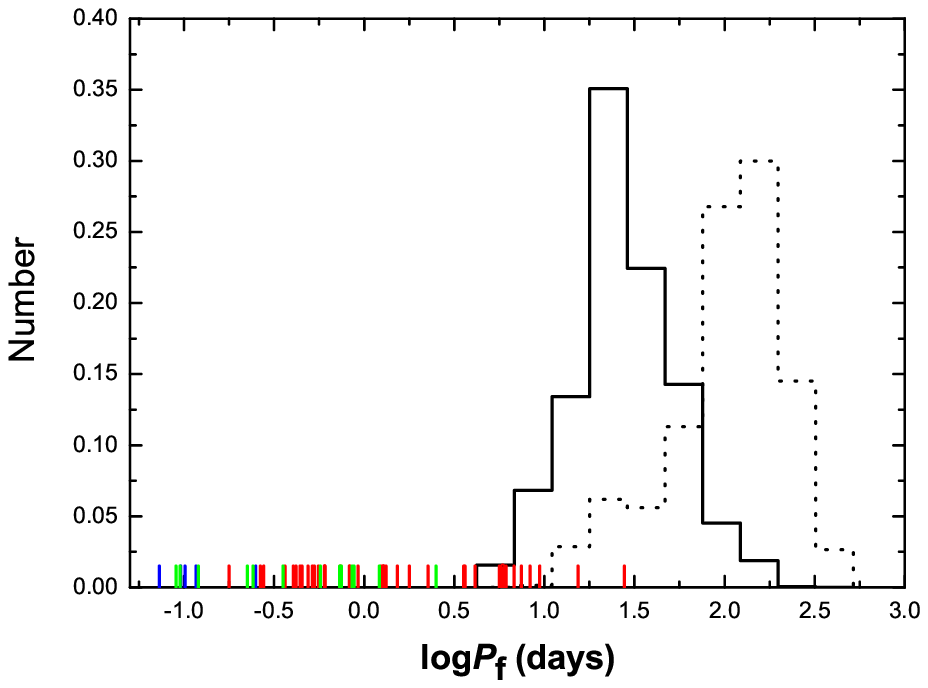}
\caption{Orbital-period distributions for stable RLOF hot subdwarfs.
The solid and dotted line indicate the conservative and
nonconservative cases, respectively. The red, blue and green bar
along the period axis indicate the orbital periods of observed
binaries with a companion of unknown type, main sequence and compact
object, respectively (Morales-Rueda et al. \cite{Morales-Rueda03},
Edelmann et al. \cite{Edelmann05}, Napiwotzki et al.
\cite{Napiwotzki04}).}
\end{figure*}

\subsection{The onset of mass transfer at different stage}

The properties of hot subdwarfs are significantly affected by the
stage where mass transfer began in their progenitors. When the onset
of mass transfer is at the main sequence, the primary would lose an
amount of mass. If its helium core is ignited, the primary would
become a hot subdwarf with a low mass and a low envelope mass. Table
3 lists the parameters of hot subdwarfs and their progenitors where
mass transfer begins at the main sequence, Hertzsprung gap or first
giant branch in the conservative case and the nonconservative case.

From our calculations, we found that the properties of the hot
subdwarfs strongly depend on the initial orbital period when $q_{\rm
i}\leq$1.5, but the dependence becomes weaker for hot subdwarfs with
high mass ratios. Figs. 10 and 13 show the Hertzsprung-Russell (HR)
diagrams and the $T_{\rm eff}$-log($g$) diagrams of hot subdwarfs
for a given initial primary mass with a different initial mass ratio
($q_{\rm i}=$1.1 and 4.0) when mass transfer begins at different
stages in the conservative and nonconservative cases. From Figs. 10
and 13, we can see that the onset of mass transfer at different
stages affects not only the surface gravity of hot subdwarfs but
also their effective temperature given $q_{\rm i}=$1.1. The surface
gravity decreases with the increase of the initial orbital period.
For $q_{\rm i}=$4.0, the onset of mass transfer at different stages
almost does not affect the effective temperature of hot subdwarfs,
but the surface gravity decreases slightly with the initial orbital
period.

In the progenitors of hot subdwarfs, the possibility of the onset of
mass transfer at the main sequence decreases with mass ratio $q_{\rm
i}$. In order to understand the mass transfer at different stages of
the primary with different initial mass, we have plotted integral
evolutionary tracks of the primary with different mass ratios in the
HR diagram and mass transfer rate in Fig. 14. From Fig. 14, we see
that the earlier the mass transfer begins, the lower the mass
transfer rate. However, mass transfer beginning at an early stage
tends to leave an sdB star. With an increase of initial orbital
period, mass transfer rates increase steadily, even up to 10$^{-4}$
$M_{\odot}$yr$^{-1}$, but the timescale of mass transfer of 10$^{5}$
yr is quite short. Some binaries with high initial mass and long
initial orbital period would be on the brink of experiencing a
delayed dynamical instability.

\begin{figure*}
\centering
\includegraphics[width=8cm,clip,angle=270]{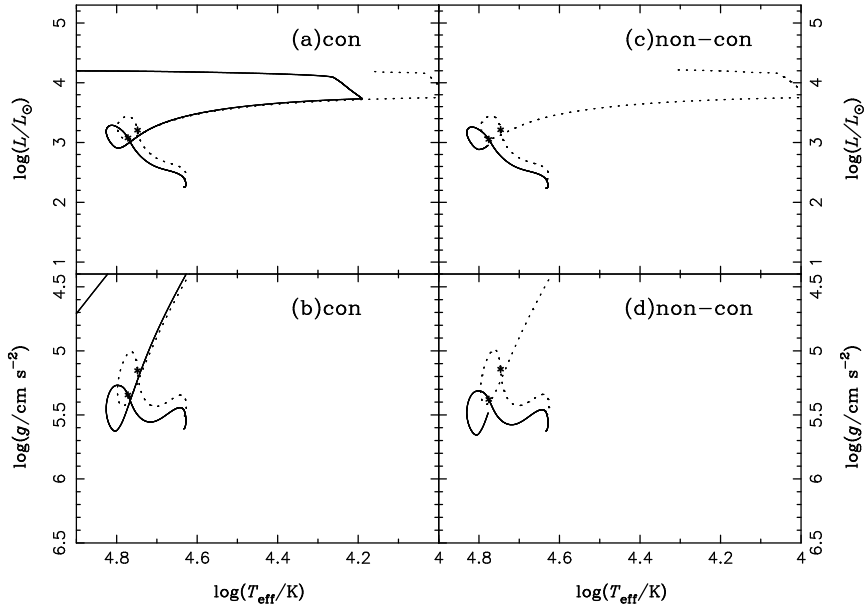}
\caption{The HR diagrams and the $T_{\rm eff}$-log$ g$ diagrams of
hot subdwarfs with progenitor$'$s initial mass $M_{\rm 1i}$ of 5.01
$M_{\odot}$ and initial mass ratio $q_{\rm i}$ is 4.0. All solid
curves are for initial orbital periods of 5.1 days (sdO stars) and
all dot-dashed for 32.3 days (sdO stars). Panels (a) and (b) are for
the conservative case. Panels (c) and (d) are for the
non-conservative case. Asterisks indicate that helium is exhausted
in its center.}
\end{figure*}

\begin{figure*}
\centering
\includegraphics[width=8cm,clip,angle=270]{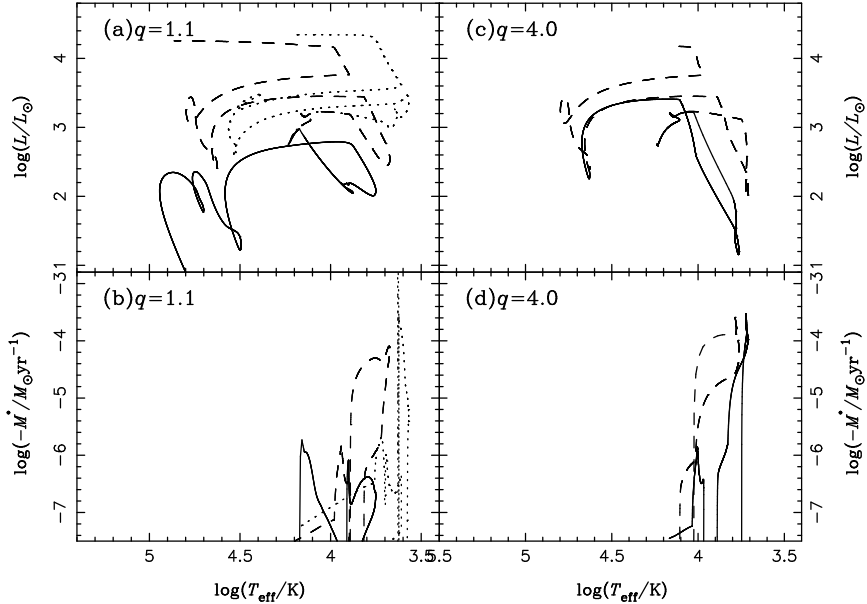}
\caption{Integral evolutionary tracks (top panels) and mass transfer
rate (bottom panels) of the primary with initial mass of $M_{\rm 1i}
= $5.01 $M_{\odot}$ in the non-conservative case. The left panels
are for $q = $1.1, and the right panels are for $q = $4.0. In the
left panels, solid curves are for the initial orbital periods of 1.6
days, mass transfer beginning at the MS stage (sdB star, with
envelope mass of 0.03 $M_{\odot}$), dashed for 15.7 days, mass
transfer beginning at the HG stage (sdO star with envelope mass of
0.09 $M_{\odot}$) and dot-dashed for 43.6 days, mass transfer
beginning at the FGB stage (sdB star with an envelope mass of 0.14
$M_{\odot}$). In the right panels, solid curves are for the initial
orbital periods of 5.1 days, mass transfer beginning at the HG stage
(sdO star with an envelope mass of 0.08 $M_{\odot}$) and dashed
curves for 32.3 days, mass transfer beginning at the HG stage (sdO
star with an envelope mass of 0.09
$M_{\odot}$).}
\end{figure*}

\begin{table}
\begin{minipage}[t]{\columnwidth}
\caption{Parameters of some typical hot subdwarfs and their
progenitors.}\footnotetext{$M_{\rm 1i} =$ initial mass; $M_{\rm f}
=$ the mass of hot subdwarfs; $q_{\rm i} =$ the mass ratio; $P_{\rm
i} =$ initial orbital period; $P_{\rm f} =$ orbital period of hot
subdwarf binaries; $M_{\rm env} =$ the envelope mass of hot
subdwarfs; $t_{\rm age} =$ the lifetime of a hot subdwarf. More
parameters can be seen from Figs. 2, 3, and 6 - 9.}
\centering
\renewcommand{\footnoterule}{}  
\begin{tabular}{lccccc}
\hline \hline $M_{\rm 1i}$ / $M_{\rm f}($\rm type$)$  & $q_{\rm i}$
& $P_{\rm i}$ / $P_{\rm f}$
 & $M_{\rm env}$  & $t_{\rm age}$ \\
($M_{\odot}$) &  & (days) & ($M_{\odot}$) & (yrs) \\
\hline
   conservative case:  \\
   3.55 / 0.38(sdB)   & 1.1  &   1.5 / 151.1   &   0.02  & 3.8$\times$10$^{8}$\\
   3.55 / 0.56(sdB)   & 1.1  &   3.5 / 116.1   &   0.04  & 0.9$\times$10$^{8}$\\
   3.55 / 0.53(sdB)   & 4.0  &   3.5 / 11.5    &   0.04  & 1.1$\times$10$^{8}$\\
   5.01 / 0.54(sdB)  & 1.1  &   1.6 / 147.8   &   0.04  & 1.0$\times$10$^{8}$\\
   5.01 / 0.65(sdOB)   & 1.1  &   2.5 / 137.3   &   0.05  & 0.6$\times$10$^{8}$\\
   5.01 / 0.93(sdO)  & 1.1  &   15.7 / 317.2   &   0.09  & 0.2$\times$10$^{8}$\\
   5.01 / 1.01(sdOB)  & 1.1  &   31.3 / 483.8   &   0.12  & 0.2$\times$10$^{8}$\\
   5.01 / 1.06(sdB)   & 1.1  &   40.6 / 545.5   &   0.16  & 0.1$\times$10$^{8}$\\
   5.01 / 0.84(sdO)   & 4.0  &   5.1 / 13.1    &   0.08  & 0.2$\times$10$^{8}$\\
   5.01 / 0.91(sdO)   & 4.0  &   32.3 / 60.8    &   0.09  & 0.2$\times$10$^{8}$\\
\hline
   non-conservative case:  \\
   3.55 / 0.33(sdB)   & 1.1   &   1.5 / 27.9   &   0.01  & 7.9$\times$10$^{8}$\\
   3.55 / 0.54(sdB)   & 1.1   &   3.5 / 29.4   &   0.04  & 1.0$\times$10$^{8}$\\
   3.55 / 0.53(sdB)   & 4.0   &   3.5 / 6.9    &   0.04  & 1.1$\times$10$^{8}$\\
   5.01 / 0.46(sdB)  & 1.1  &   1.6 / 27.2   &   0.03  & 1.7$\times$10$^{8}$\\
   5.01 / 0.61(sdOB)   & 1.1  &   2.5 / 29.0   &   0.05  & 0.7$\times$10$^{8}$\\
   5.01 / 0.91(sdO)  & 1.1  &   15.7 / 100.5   &   0.09  & 0.2$\times$10$^{8}$\\
   5.01 / 0.98(sdOB)  & 1.1  &   41.6 / 483.2   &   0.12  & 0.2$\times$10$^{8}$\\
   5.01 / 1.00(sdB)   & 1.1  &   43.6 / 526.0   &   0.14  & 0.1$\times$10$^{8}$\\
   5.01 / 0.84(sdO)   & 4.0  &   5.1 / 9.1    &   0.08  & 0.3$\times$10$^{8}$\\
   5.01 / 0.91(sdO)   & 4.0  &   32.3 / 59.4    &   0.09  & 0.2$\times$10$^{8}$\\
\hline
\end{tabular}
\end{minipage}
\end{table}

\subsection{Birth rate and number of hot subdwarfs from the stable RLOF channel}

Table 4 lists the birth rates of hot subdwarfs formed in various
formation channels. In the table, the first column denotes the
metallicity ($Z$=0.02 for Population I); the second column gives
$q_{\rm crit}$, the critical mass ratio for the first stable RLOF on
the FGB; Columns 3-8 list Galactic birth rates for hot subdwarfs (in
10$^{-3}$ yr$^{-1}$) from the stable RLOF channel of
intermediate-mass binaries (our model), the first stable RLOF
channel, the first CE ejection channel, the second RLOF channel, the
second CE ejection channel and the helium WD merger channel.

As Table 4 shows, the predicted birth rate of Population I hot
subdwarfs from the stable RLOF channel of intermediate-mass binaries
is 0.004 yr$^{-1}$ for the Galaxy. By taking an effective Galactic
volume of 5 $\times$ 10$^{11}$ pc$^{3}$ (Zombeck \cite{Zombeck90}),
we give an average birth rate per pc$^{3}$ of 8 $\times$ 10$^{-15}$
pc$^{-3}$ yr$^{-1}$. When convolved with the lifetime of hot
subdwarf phase, this rate implies a total number of hot subdwarfs
from the stable RLOF channel of intermediate-mass binaries in the
Galaxy of 0.8 $\times$ 10$^{6}$. Comparing this result with the
result of Han et al. (\cite{Han03}), we conclude that the stable
RLOF channel of intermediate-mass binaries is important.

In order to compare our results with observed hot subdwarfs in the
$T_{\rm eff}$-log($g$) diagram, we plot a grey diagram as Fig. 15.
Comparing this figure with Fig. 4, we suppose that some hot
subdwarfs with $T_{\rm eff}$ $>$ 30,000 K and log($g$)$>$ 6.0 might
be evolved sdB stars.

\begin{table}
\begin{minipage}[t]{\columnwidth}
\caption{Birth rates of hot subdwarfs from the stable RLOF channel
of intermediate-mass binaries (our model) and other channels (Han et
al. \cite{Han03}) (in 10$^{-3}$ yr$^{-1}$).}
\footnotetext{$Z$=metallicity; $q_{crit}$=the critical mass ratio for
the first stable RLOF on the FGB; $\nu_{IMR}$=Galactic birth rates
of hot subdwarfs from the stable RLOF channel of intermediate-mass
binaries; $\nu_{FR}$=the first stable RLOF channel; $\nu_{FCE}$=the
first CE ejection channel; $\nu_{SR}$=the second RLOF channel;
$\nu_{SCE}$=the second CE ejection channel; $\nu_{M}$=the helium WD
merger channel.}
 \centering
\renewcommand{\footnoterule}{}  
\begin{tabular}{lcccccccc}
 \hline
 \hline
 \\
 $Z$  &  $q_{crit}$  &  $\nu_{IMR}$  & $\nu_{FR}$
 & $\nu_{FCE}$ & $\nu_{SR}$ & $\nu_{SCE}$ & $\nu_{M}$\\
 \\
\hline
\\
0.02   &  1.5  &  4.18  &   29.89& 8.41 &0.00&8.38&17.22\\
\\
0.02  &  1.2  &  4.14  &   22.25 &10.71 &0.00&5.43&7.99\\
\\
\hline
\end{tabular}
\end{minipage}
\end{table}

\begin{figure*}
\centering
\includegraphics[width=10cm,clip,angle=0]{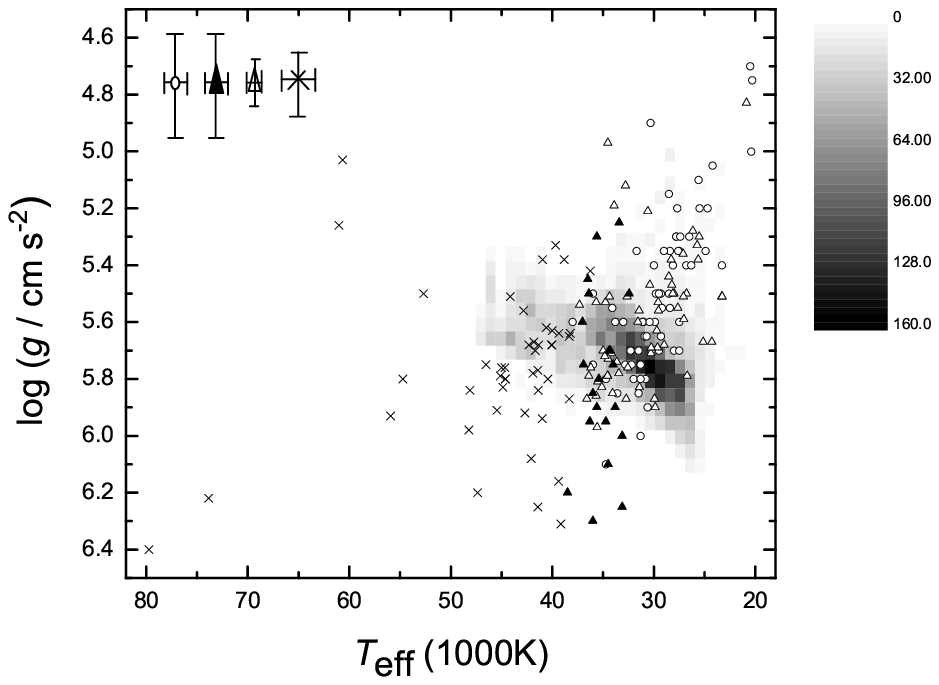}
\caption{Comparison of our results to the atmospheric parameters of
sdB stars (open circles Edelmann et al. \cite{Edelmann03}, open
triangles Lisker et al. \cite{Lisker05}), sdOB stars (solid
triangles Edelmann et al. \cite{Edelmann03}), sdO stars (crosses
Stroeer et al. \cite{Stroeer07}). The largest cross is for HE
1115-0631 (orbital period: 5.87 d, Napiwotzki et al.
\cite{Napiwotzki04}). Our results are shown as shaded $T_{\rm
eff}$-log($g$)-boxes, where a higher hot subdwarf density per box
corresponds to darker shading. The grey scale is shown in the right
of the figure.}
\end{figure*}

\subsection{The effect of the secondary}

Most hot subdwarfs contain a main sequence (MS) or a subgiant star
in our models. We discuss the effect of the secondary on
observations for hot subdwarfs with a companion MS star in this
section.

The luminosity of the secondary ($Z=$0.02) is fitted by Tout et al.
(\cite{Tout96}) as:
\begin{equation}
L=\frac{\alpha M^{5.5}+\beta M^{11}}{\gamma+M^{3}+\delta
M^{5}+\varepsilon M^{7}+\zeta M^{8}+\eta M^{9.5}},
\end{equation}
where $M$ is the mass of a main sequence star, and $\alpha, \beta,
\gamma, \delta, \varepsilon, \zeta, \eta$ are constants (see Tout et
al. \cite{Tout96}). Our calculation indicates that if $q_{\rm
i}\leq$2.0, the luminosity of almost all hot subdwarfs is lower than
that of their companions. This implies that these hot subdwarfs
would not be detected by ground-based telescopes, although they
could be found by space telescopes. If $q_{\rm i}\geq$3.0, the
luminosity of most hot subdwarfs will be larger than that of their
companions, and this indicates that these hot subdwarfs might be
observed by ground-based telescope. So, we suggest that a small
fraction of observed hot subdwarfs could have progenitors with
$q_{\rm i}$ larger than $\sim$3.0.

\section{Discussion}

Han et al. (\cite{Han02}) have proposed a binary model for the
formation of hot subdwarfs, including three channels: the CE
ejection channel, the stable RLOF channel and the double helium WD
merger channel. In the CE ejection channel, a primary experiences
dynamically unstable mass transfer at the tip of the FGB and leads
to the formation of a CE (Paczy$\acute{\rm n}$ski
\cite{Paczynski76}). During the spiral-in process in the CE, the
orbit of the binary shrinks due to the friction between binary and
envelope. The CE will be ejected if the released orbital energy
exceeds the binding energy in this process, and this process would
leave a very tight binary containing the degenerate core of the
giant. If the degenerate core experiences a helium flash, the
remnant core of the giant may be ignited (Castellani \& Castellani
\cite{Castellani93}) and hence appear as a core-helium-burning hot
subdwarf in binaries with short orbital periods in the range of 0.05
- 40 days (Han et al. \cite{Han02}). In stable RLOF channel, a
primary undergoes a stable mass transfer at the FGB stage. If its
helium core was ignited after the RLOF, the primary would become a
hot subdwarf in binaries with long orbital periods in the range of
400 - 1500 days (Han et al. \cite{Han02}). Our results indicate that
the mass transfer can also begin at the MS stage (case A mass
transfer) and HG stage (case B mass transfer). This leads to hot
subdwarfs in binaries with a wide orbital period range of 5 - 900
days. Compared with the CE ejection channel, we found that there are
a large number of hot subdwarfs in binaries with long orbital
periods formed through the stable RLOF channel. This would sustain
the observations by Green, Liebert \& Saffer (\cite{Green00}) and
Morales-Rueda et al. (\cite{Morales-Rueda03}), who argued that some
sdB stars could appear to be members of long period binaries.
Moreover, it is possible that some hot subdwarfs in binaries with a
very long orbital period could be misunderstood as single stars.

In general, sdB stars are believed to have a mass of around 0.5
$M_{\odot}$ and a thin hydrogen rich envelope around 0.02
$M_{\odot}$ (Heber \cite{Heber86}, Saffer et al. \cite{Saffer94}).
Han et al. (\cite{Han02}) suggested that the mass of sdB stars is in
the range of 0.33 - 0.68 $M_{\odot}$. In our models, the mass of
most sdB stars is in the range of 0.32 - 0.67 $M_{\odot}$ with a
thin envelope of 0.01 - 0.05 $M_{\odot}$, but a few sdB stars have
an anomalously high mass of 0.72 - 1.13 $M_{\odot}$, and a thick
envelope of 0.07 - 0.16 $M_{\odot}$. The reasons have been described
in section 3. The mass of sdO stars is from 0.75 $M_{\odot}$ to 1.44
$M_{\odot}$ and its envelope mass is higher than 0.07 $M_{\odot}$,
even up to 0.22 $M_{\odot}$. The mass and the envelope mass of sdOB
stars are between sdB stars and sdO stars. In addition, Han et al.
(\cite{Han00}) and Chen et al. (\cite{Chen02}) discussed low- and
intermediate-mass close binary evolution and the initial-final mass
relation of these binaries. By calculating a binary evolution
sequence with an initial mass ratio between 1.1 and 4.0, and the
onset of the stable RLOF at the Hertzsprung gap stage, they found
that the remnant mass of the primary depends on the initial mass
ratio or orbital period for a given primary mass. These results are
similar to ours, but binary evolution calculations with the onset of
mass transfer at the main sequence stage have been included in our
models.

Elson et al. (\cite{Elson98}) discovered a blue star with $T_{\rm
eff}\approx$ 31,500 K, log($g$)$\approx$ 4.4, log$L/L_{\odot}~\sim~$
3.0 (Burleigh et al. \cite{Burleigh99}) in the young cluster NGC
1818 in the Large Magellanic Cloud (LMC), which has a main-sequence
turnoff mass of $\sim$7.5 - 9.5 $M_{\odot}$, an age of $\sim$2 - 4
$\times$ 10$^{7}$ yr (Will et al. \cite{Will95}) and a metallicity
of $\sim$0.005 (Kerber \& Santiago \cite{Kerber05}). This star was
considered as a candidate of luminous WD. However, Liebert
(\cite{Liebert99}) and Burleigh et al. (\cite{Burleigh99}) pointed
out that the star would be a post-EHB star or EHB star, rather than
a WD, and it would lie in the Galactic halo or the disc of the LMC
instead of cluster NGC 1818. If this blue star was in a binary
system, we suggest that the star could be in the cluster NGC1818 and
form through the stable RLOF channel of intermediate-mass binaries,
but this needs more observational and theoretical evidence to
confirm. We consider that hot subdwarfs could exist in young
clusters, and some of them might have high mass ($>$ 0.7
$M_{\odot}$). We will focus on hot subdwarfs in young clusters and
calculate a binary evolution sequence with different metallicity
($Z$) to study their properties, formation and evolution in the
future.

Mass transfer efficiency is an important factor affecting the
initial parameter spaces, as lower mass transfer efficiency will
lead to more binaries experiencing stable RLOF rather than the
formation of a common envelope during rapid mass transfer. Past work
(Paczy$\acute{\rm n}$ski \& Zi$\acute{\rm o}\l$kowski
\cite{Paczynski67}; Refsdal, Roth \& Weigert \cite{Refsdal74}; De
Greve \& De Loore \cite{deGreve92}; Chen \& Han
\cite{Chen02,Chen03}) employed mass transfer efficiency $\beta~=$
0.5 for the calculation of stellar evolution in nonconservative
cases. However, De Mink et al. (\cite{deMink07}) suggested that mass
transfer efficiency should not be a single constant by comparing
their models with a sample of 50 double-lined eclipsing binaries in
the Small Magellanic Cloud, which could be affected by spin up of
the accreting star and tidal interaction. They also found that
initially wider systems tend to favor less conservative models,
since accreting angular momentum will speed up the rotation of the
accreting star, resulting in mass loss along its equator region, as
discussed by Wellstein (\cite{Wellstein01}) and Petrovic et al.
(\cite{Petrovic05}). Our results are in line with this conclusion.
Due to the uncertain of the mass transfer efficiency (De Mink et al.
\cite{deMink07}), the birth rate and the total number from our
models need further data to constrain. However, it is reasonable to
estimate that a high birth rate and large total number of hot
subdwarfs via stable RLOF channel will suggest less conservative
models.

Other effects on the birth rate and total number of hot subdwarfs in
our models would arise from the life-time of the secondary, which
will become too short, such that it is comparable to or even smaller
than the lifetime of the hot subdwarf phase (typical value:
2.0$\times$10$^{8}$ yr) in some extreme cases. In these two cases,
the secondary presumably evolved rapidly to the giant branch or even
to later stages after fast mass transfer, while the primary could be
a giant or a hot subdwarf or a helium star, which could be
associated with a $'$post-Algol$'$ binary (Nelson \& Eggleton
\cite{Nelson01}). As a result, we could overestimate the birth rate
and the total number of hot subdwarfs via the stable RLOF channel.

\section{Conclusions}

We present the properties of hot subdwarfs from the stable RLOF
channel of intermediate-mass binaries by computing population I
binary evolution sequences both in conservative and nonconservative
cases, where the primary has an initial mass in the range of 2.2 -
6.3 $M_{\odot}$, initial mass ratio $q_{\rm i}=$1.1, 1.5, 2.0, 3.0,
4.0, 4.2, 4.5, and onset of RLOF at the MS, HG, and FGB stage. Due
to the effect of the secondary, the ratio of the hot subdwarf
bianries which might be observed by ground-based telescope would be
larger around 3.0. We summarize our results as follows:
   \begin{enumerate}

\item The birth rate of hot subdwarfs from the stable RLOF channel of
intermediate-mass binaries is $\sim$4.2$\times10^{-3}$ yr$^{-1}$,
which is similar to the first CE ejection channel. However, the
current observations indicate that only a few hot subdwarfs binaries
could be formed via the stable RLOF channel (Fig. 12). This could be
associated with the rotation of the accreting star and the evolution
of the secondary.

\item We have obtained the initial parameter space of the progenitor binaries
of hot subdwarfs from the stable RLOF channel of intermediate-mass
binaries (Figs. 2 and 3).

\item  Our results indicate that the hot subdwarfs from the stable RLOF channel of
intermediate-mass binaries have a mass in the range of 0.32 - 1.44
$M_{\odot}$ with a wide envelope mass of 0.01 - 0.22 $M_{\odot}$.
Most sdB stars have a mass in the range of 0.32 - 0.67 $M_{\odot}$
with an envelope mass of 0.01 - 0.05 $M_{\odot}$. Most sdOB stars
have a mass in the range of 0.60 - 0.86 $M_{\odot}$ with an envelope
mass of 0.05 - 0.10 $M_{\odot}$. SdO stars have a mass in the range
of 0.75 - 1.44 $M_{\odot}$ with an envelope mass of 0.07 - 0.22
$M_{\odot}$. Furthermore, we found that some sdB or sdOB stars have
anomalously high mass, in the range of 0.72 - 1.13 $M_{\odot}$ and
0.97 - 1.23 $M_{\odot}$, respectively. The envelope mass is in the
range of 0.07 - 0.16 $M_{\odot}$ and 0.11 - 0.19 $M_{\odot}$,
respectively. We expect that observations could confirm these
theoretical results in the future.

\item The orbital periods of the binaries have a
wide range of 5.0 - 900.0 days, while the peak is at around 120 days
for the conservative case and 30 days for the nonconservative case.

\item Our results favor that hot subdwarfs, formed through the RLOF
channel of intermediate-mass binaries, will be found in young
clusters in the future.

   \end{enumerate}

\begin{acknowledgements}
We would like to acknowledge Zhanwen Han, Fenghui Zhang and Xuefei
Chen for their discussions and suggestions and also thank the
referee and editor for their useful suggestions and comments.
\end{acknowledgements}

\end{document}